\documentclass[doublecol,linenumbers]{epl2}
\usepackage{graphicx}
\usepackage{amsmath}
\usepackage{makeidx}
\usepackage{xcolor}
\usepackage{amssymb}
\usepackage{float}

\title{Wave packet dynamics for a non-linear Schrodinger equation: Qualitative changes with changes in the initial width}
\author{Sukla Pal\inst{1} \and {J. K. Bhattacharjee}\inst{2}}
\institute{
\inst{1}Department of Theoretical Physics, S.N.Bose National Centre For Basic Sciences, JD-Block, Sector-III, Salt Lake City, Kolkata-700098, India\\
\inst{2}Harish-Chandra Research Institute, Chhatnag road, Jhunsi, Allahabad-211019, India
}
\date{Received: date / Revised version: date}

\abstract{The propagation of an initially Gaussian wave packet of width $\Delta_0$ in a cubic non-linear Schrodinger equation with a negative coupling constant for the nonlinear term is considered . It is predicted analytically and verified numerically that for a free particle if $\Delta_0$ is less than a critical value $\Delta_c$, then the packet will propagate in time with linearly growing width but for $\Delta>\Delta_c$, the packet will start becoming narrow and cease to be a Gaussian . For a simple harmonic oscillator, we find that for $\Delta_0$ smaller than a critical value, there always exist a coupling strength for which the packet simply oscillates about the mean position without changing its shape.}

\pacs{05.}{Nonlinear dynamical systems} 
\begin{document}
\maketitle
The dynamics of wave packets has been well studied in literature both in case of free particle and harmonic oscillator in quantum mechanics. Among all the wave packets the Gaussian wave packet has always been the most popular due to its calculational simplicity and well understood nature. Here we want to discuss the propagation of an initial Gaussian wave packet in a Gross-Pitaevskii equation (GPE) when the potential is absent (making it a nonlinear Schrodinger equation (NLSE)) and when the potential is simple harmonic. The usual dynamics that has been discussed for the NLSE is the formation of solitons. For positive non-linearity (repulsive interparticle interaction) there appears dark solitons (density dip) and if the nonlinear term becomes negative the formation of bright soliton (density peak)has been observed \cite{l}-\cite{r}. We take a different course here and study the dynamics of an initial Gaussian wave packet, both for NLSE and GPE with a simple harmonic potential. We have two primary results:\\

i) For the NLSE with a negative non-linear term, the Gaussian wave packet simply translates with an ever increasing width if the initial wave packet has a width which is smaller than a critical width. For widths greater than the critical width on the other hand, the wave packet becomes narrow and the Gaussian shape changes dramatically.

ii) In a GPE with a harmonic trap, an initial Gaussian wave packet will simply oscillate with a fixed width if \\
a)the non linear term in $\psi$ has a negative strength\\
b)the dimensionless width $\delta$ satisfy $\delta^2<\sqrt{\frac{1}{3}}$\\
c)the strength of the nonlinear term is a given function of $\delta$\\

We will arrive at the critical values of the parameters relevant for both i) and ii) analytically and then numerically establish the truth of the above assertions.

In one spatial dimension, GPE takes the following form
\begin{equation}\label{eq1}
i\hbar\frac{\partial\psi}{\partial t}=-\frac{\hbar^2}{2m}\frac{\partial^2\psi}{\partial x^2}+V(x)\psi+g|\psi|^2\psi
\end{equation}
where, $V(x)$ is an external potential. If $V(x)=0$, then then the equation is generally known as the nonlinear Schrodinger equation (NLSE). To begin with, we consider the equation of motion for the expectation value of any operator (say O) with $\psi$ governed by the Eq.(\ref{eq1}) which can be written as follows
\begin{eqnarray}\label{eq2}
\frac{d}{dt}\langle O\rangle=\frac{1}{i\hbar}\langle[O,H]\rangle+\frac{g}{i\hbar}\int\psi^{\ast}(O|\psi|^2-|\psi|^2O)\psi dx
\end{eqnarray}
where, $H=-\frac{\hbar^2}{2m}\frac{\partial^2}{\partial x^2}+V(x)$ will be called the Hamiltonian of the concerned system. The time evolution of the expectation values obey the equations given below:
\begin{eqnarray}
\frac{d\langle x\rangle}{dt} &=& \frac{\langle p\rangle}{m}\label{eq3}\\
\frac{d\langle x^2\rangle}{dt} &=& \frac{1}{m}\langle xp+px\rangle\label{eq4}\\
\frac{d\langle p\rangle}{dt} &=& -\langle\frac{dV}{dx}\rangle\label{eq5}\\
\frac{d\langle p^2\rangle}{dt} &=& -\langle p\frac{dV}{dx}+\frac{dV}{dx}p\rangle- mg\frac{d}{dt}\int|\psi|^4dx\label{eq6}\\
\frac{d\langle xp+px\rangle}{dt} &=& 4[\frac{\langle p^2\rangle}{2m}-\frac{1}{2}\langle x\frac{dV}{dx}\rangle]+g\int|\psi|^4dx\label{eq7}
\end{eqnarray}
We treat the case of free particle and simple harmonic oscillator separately.
\section{Free particle}
For the free particle where $V(x)=0$, we find $\langle p\rangle=p_0$ (constant) and $\langle x\rangle=\frac{p_0}{m}t+a$, where $p_0$ and $x_0$ are the initial value of $\langle p\rangle$ and $\langle x\rangle$ respectively. We take an initial Gaussian wave packet 
\begin{equation}\label{eq28}
\psi(x,t=0)=\frac{1}{\pi^{\frac{1}{4}}\Delta_0^{\frac{1}{2}}}e^{-\frac{(x-a)^2}{2\Delta_0^2}}e^{i\frac{p_0}{\hbar}x}
\end{equation}
If $\Delta^2=\langle x^2\rangle-\langle x\rangle^2$, then the above equations lead to 
\begin{equation}\label{eq31}
\frac{d^2}{dt^2}\Delta^2=\frac{g}{m\sqrt{2\pi}}\frac{1}{\Delta}+\frac{\hbar^2}{2m^2\Delta_0^2}
\end{equation}
In the first term of Eq.(\ref{eq31}), we have assumed the shape to remain in Gaussian. As is seen from the definition, $\Delta$ is the width of the packet. Writing $\Delta^2=Y$, this dynamics turns out to be 
\begin{equation*}
\ddot{Y}=-\frac{\partial V}{\partial Y}
\end{equation*}
where the potential $V$ is given by
\begin{equation}\label{eq31a}
V=-\frac{\hbar^2}{2m^2\Delta_0^2}Y-\frac{2g}{m\sqrt{2\pi}}Y^{1/2}
\end{equation}
\begin{figure}[H]
\includegraphics[angle=0,scale=0.9]{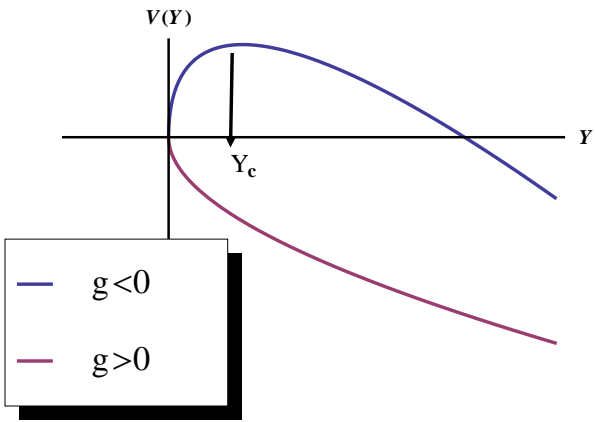}
\caption{This figure shows effective potential $ V(Y)$ with $Y$ for free particle wave packet. For $g>0$ the curve shows linear fall with $Y$ indicating absence of any equilibrium width. For $g<0$, there is a maximum at $Y_c$ which leads to drastically different dynamics in either side of $Y_c$.}
\label{fig4}
\end{figure}
The nature of the effective potential has been shown in Fig.(\ref{fig4}) both for positive and negative $g$. For $g>0$, we have an increasing $\Delta(t)$ for all values of $\Delta_0$. However if $g<0$, then the effective potential has an extremum. The maximum of $V(Y)$ occurs at $Y_c=[\frac{2m}{\hbar^2}(\frac{-g}{\sqrt{2\pi}})\Delta_0^2]^2$.If the initial width $\Delta_0 $ is such that $Y_0=\Delta_0^2<Y_c$, then the maximum possible value of $Y$ is limited by $Y_c$ and as time goes on $\Delta$ decreases and becomes zero at a finite time. The condition for this is $\Delta_0>\Delta_c=\frac{\hbar^2}{m}(\frac{\sqrt{2\pi}}{2|g|})$. i.e., the initial width has to be greater than a critical value. The fact that $\Delta$is driven to zero at a finite time indicates the wave packet will not remain Gaussian after a finite time interval. On the other hand if $\Delta_0<\Delta_c$, then the wave packet width will increase with time becoming a linear increase at later times. The dependence of $\Delta_c$ on $|g|$ shows that for large $g$, even an initially narrow packet can show this unexpected behavior. 

To make our theory dimensionless we fix our length and time scale by $\frac{\hbar}{p_0}$ and $\frac{m\hbar}{p_0^2}$ respectively. Also we consider $\frac{g}{\sqrt{2\pi}}=\gamma(\frac{p_0^2}{m})^2\frac{\hbar}{p_0}$ such that $\gamma$ turns out to be the dimensionless coupling constant of the theory. After considering the above scaling we rewrite Eq.(\ref{eq31}) in the following dimensionless form ($\bar{\Delta}$ is dimensionless)
\begin{equation}\label{eq32}
\frac{d^2}{dt^2}\bar{\Delta}^2=\gamma\frac{1}{\bar{\Delta}}+\frac{1}{2\bar{\Delta}_0^2}
\end{equation}
 and hence following the previous argument we find that for $g<0$, the initial width $\bar{\Delta}_0$ satisfies $\bar{\Delta}_0>\frac{1}{2\gamma}$ for the packet to spread in time. From now on we will consider $\bar{\Delta}\rightarrow\Delta$.
\begin{figure}[H]
\includegraphics[angle=0,scale=0.95]{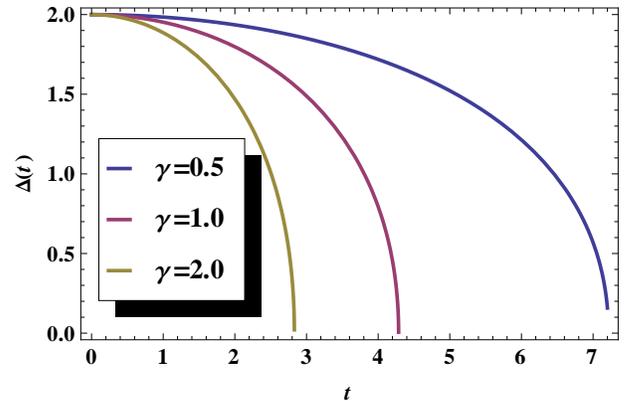}
\caption{The variation of free particle wave packet width $\Delta(t)$ with time $t$ for $\Delta_0>\frac{1}{2\gamma}(=Y_c)$ with $\gamma<0$. The plots are shown for three different values of $\gamma$ ($\gamma$=-0.5, -1.0, -2.0) and for $\Delta_0=2.0$}
\label{fig5}
\end{figure}
\begin{figure}[H]
\includegraphics[angle=0,scale=0.95]{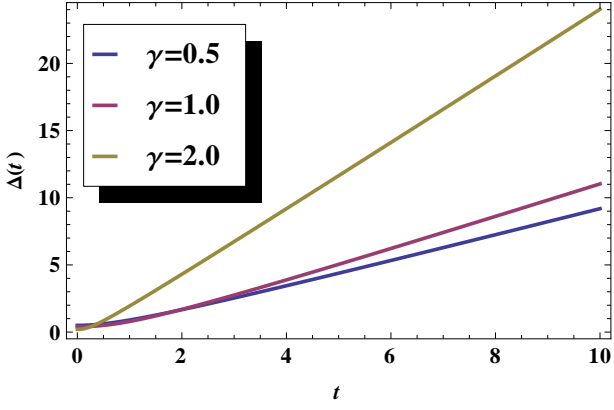}
\caption{Time evolution of free particle wave packet width $\Delta(t)$ for $\Delta_0<\frac{1}{2\gamma}(=Y_c)$ with $\gamma<0$. The width of the wave packet spreads rapidly over time as is expected for free particle. The blue curve is plotted for $\gamma=-0.5$ and $\Delta_0=0.5$; violet is for $\gamma=1.0$ and $\Delta_0=0.4$; and dark yellow is for $\gamma=2.0$ and $\Delta_0=0.2$. }
\label{fig6}
\end{figure}
\section{Simple Harmonic Oscillator}
We take the simple harmonic potential $V(x)=\frac{1}{2}m\omega^2x^2$ and consider the initial Gaussian wave packet to remain a Gaussian of width $\Delta(t)$ and centered at $x_0(t)$ moving to the right with momentum $p(t)$, such that
\begin{equation}\label{eq15}
\psi(x,t)=\frac{1}{\pi^{\frac{1}{4}}\Delta^{\frac{1}{2}}}e^{-\frac{(x-x_0(t))^2}{2\Delta^2(t)}}e^{i\frac{p(t)}{\hbar}x}
\end{equation}
Clearly for Gaussian wave packet $\langle[x-x_0(t)]^2\rangle=\Delta^2(t)$, $\langle p\rangle=p(t)$ and $\langle x\rangle=x_0(t)$. We set initial condition with $x_0(t=0)=a$, $p(t=0)=0$ and $\Delta(t=0)=\Delta_0$ so that the initial wave packet has the form like $\psi(x,0)=\frac{1}{\pi^{\frac{1}{4}}\Delta_0^{\frac{1}{2}}}e^{-\frac{(x-a)^2}{2\Delta_0^2}}$. Considering this initial Gaussian wave packet, few simple calculation leads to the following dynamical equation for the wave packet width.
\begin{equation}\label{eq19}
\frac{d^2\Delta^2}{dt^2}=D-4\omega^2\Delta^2-\frac{g}{m\sqrt{2\pi}}\frac{1}{\Delta}
\end{equation}
with constant $D$ given by
\begin{equation}\label{eq20}
D=\frac{\hbar^2}{m^2\Delta_0^2}+\omega^2\Delta_0^2+\frac{2g}{m\sqrt{2\pi}}\frac{1}{\Delta_0}
\end{equation}
Defining $\Delta^2=Y$, the equation of motion now takes the following simple form
\begin{equation}\label{eq21}
\ddot{Y}=D-4\omega^2Y-\frac{g}{m\sqrt{2\pi}}\frac{1}{\sqrt{Y}}
\end{equation}
Dynamics of $Y$ governed by Eq.(\ref{eq21}) is equivalent to the motion in the effective potential given below
\begin{equation}\label{eq22}
V(Y)=2\omega^2Y^2+\frac{2g}{m\sqrt{2\pi}}Y^{1/2}-DY
\end{equation}
It is obvious from Eq. (\ref{eq20}) that for $g>0$, $ D$ can never take negative values. For both $g>0$ and $D>0$, the potential can have extrema satisfying the following constraint.
\begin{equation}\label{eq23}
4\omega^2Y_c+\frac{g}{2m\sqrt{2\pi}}\frac{1}{Y_c^{1/2}}=D
\end{equation}
Eq.(\ref{eq23}) dictates that the extrema can only exist if $D>3(4\omega^2)^{1/3}(\frac{g}{4m\sqrt{2\pi}})^{2/3}$.\\

The interesting situation appears when $g<0$. The effective potential now takes the form $V(Y)=2\omega^2Y^2-DY-\frac{|g|}{m\sqrt{1\pi}}Y^{1/2}$. The potential has one minimum and it is located by $Y_{min}$, where
\begin{equation}\label{eq24}
\frac{|g|}{m\sqrt{2\pi}}\frac{1}{Y_{min}^{1/2}}=4\omega^2Y_{min}-D
\end{equation}
We fix the initial width at $\Delta_0=\delta\sqrt{\frac{\hbar}{m\omega}}$ where $\delta$ is a number of O(1). Since $\sqrt{\frac{\hbar}{m\omega}}$ is the characteristic length scale of the oscillator, we use it to define the dimensionless shift $\alpha$ of the peak of the wave packet and dimensionless coupling constant $\beta$ as $a=\sqrt{\frac{\hbar}{m\omega}}\alpha$ and $\frac{g}{\sqrt{2\pi}}=\beta\hbar\omega\sqrt{\frac{\hbar}{m\omega}}$.
Such that the $D$ in Eq.(\ref{eq20}) can be written as 
\begin{equation}\label{eq26}
D=\frac{\hbar\omega}{m}[\delta^2+\frac{1}{\delta^2}-2\frac{|\beta|}{\delta}]
\end{equation}
By using above scalings, the effective potential takes the following dimensionless form 
\begin{equation*}
V=2Y^2-\beta\sqrt{Y}-(\delta^2+\frac{1}{\delta^2}-2\frac{|\beta|}{\delta})Y
\end{equation*}
where V is in the scale of $\omega^2(\frac{\hbar}{m\omega})^2$.
\begin{figure}[H]
\includegraphics[angle=0,scale=0.95]{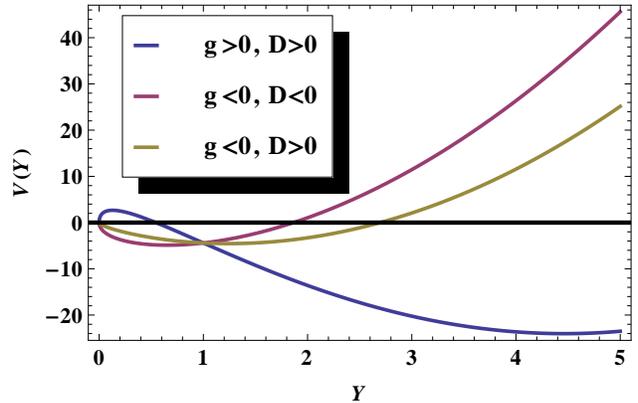}
\caption{The plot of Effective potential $ V(Y)$ with $Y$ for SHO. For $g>0$ the curve shows in this case an extrema. For all values of $g<0$, there exist a minima in the potential indicating $Y_{min}$.}
\label{fig1}
\end{figure}
Returning to the dynamics of $Y(=\Delta^2)$, we note that for the initial wave function of Eq.(\ref{eq15})the initial value of $\frac{d\Delta}{dt}$, obtained from Eq.(\ref{eq4}, is clearly zero. Hence with $\frac{d\Delta}{dt}=0$ at $t=0$, if $\Delta(t=0)$ is set at $Y_{min}$, then the width will remain at $Y_{min}$ forever and we will have a wave packet that will oscillate with this frequency provided the initial wave packet has a non zero wave number. Since for the calculation of $D$ in Eq.(\ref{eq26}), we have set $\Delta_0=\delta\sqrt{\frac{\hbar}{m\omega}}$, this requires $Y_{min}=\Delta_0^2=\delta^2\frac{\hbar}{m\omega}$ and hence Eq.(\ref{eq24}) becomes
\begin{eqnarray}\label{eq27}
(\frac{1}{\delta^2}-3\delta^2)=\frac{|\beta|}{\delta}
\end{eqnarray}
If $\delta<(\frac{1}{3})^{1/4}$ there will always exist a value of $\beta$ for which the initial wave packet simply oscillates with almost no change of width. In dimensionless unit Eq.(\ref{eq21}) takes the following simple form, where all the lengths are in the scale of harmonic oscillator length and time is in the scale of inverse trapping frequency $\omega^{-1}$.
\begin{equation}\label{eq27a}
\ddot{Y}=\delta^2+\frac{1}{\delta^2}-2\frac{|\beta|}{\delta}-4Y+\frac{|\beta|}{\sqrt{Y}}
\end{equation}
\begin{figure}[!htbp]
\includegraphics[angle=0,scale=0.9]{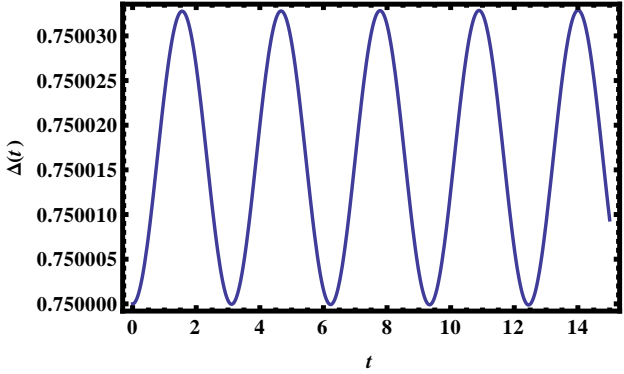}
\caption{$\Delta(t)$ vs. $t$ when $\delta=0.75$ is considered and Eq.(\ref{eq27}) is satisfied. }
\label{fig3}
\end{figure}
Fig.\ref{fig3} clearly agrees with the initial condition $(\delta<(\frac{1}{3})^{1/4})$ as we predicted analytically for existence of constant width wave packet, observed oscillatory time dependence of $\Delta(t)$ being negligibly small.
\begin{figure}[H]
\includegraphics[angle=0,scale=0.9]{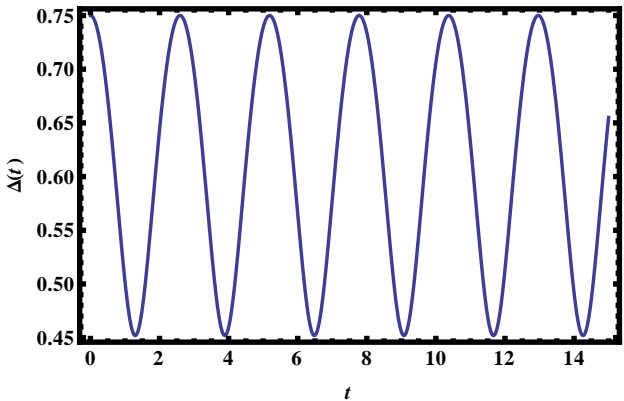}
\caption{$\Delta(t)$ vs. $t$ when $\delta=0.75$ is considered but when Eq.(\ref{eq27}) is not satisfied. $\beta=0.8$ is considered. The width oscillates in time significantly. }
\label{fig7}
\end{figure}
A linear stability analysis of Eq.(\ref{eq27a}) about $Y_{min}$, gives an oscillation about $Y_{min}$ with a frequency of $\sqrt{4+\frac{|\beta|}{2Y_{min}^{3/2}}}$ which in this case gives $\frac{\omega}{2\pi}=0.32$. From Fig.\ref{fig7}, the frequency turns out to be 0.4. Thus the slight perturbation from $Y_{min}$ will make system oscillate around it with the above frequency. The frequency obtained from linear stability is quite close to the actually observed one.
\section{Numerical Study}
\begin{figure}[H]
\includegraphics[angle=0,scale=0.8]{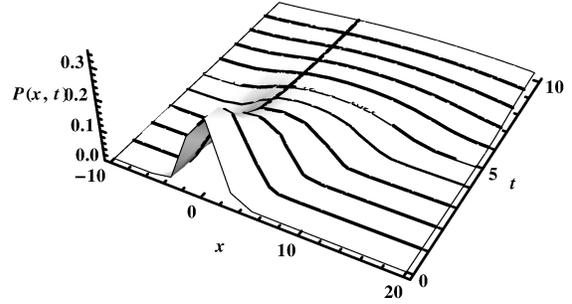}
\caption{Time evolution of initial Gaussian wave packet for $\gamma>0$. ($\gamma=2.0$ and $\delta=2.0$)for a free particle. With time the initial wave packet spreads over the space.}
\label{free-dyn-1}
\end{figure}

For free particle, considering the following transformations: $t'=\frac{p_0^2}{m\hbar}t$, $x'=\frac{p_0}{\hbar}x$,$\psi'=\sqrt{\frac{\hbar}{p_0}}\psi$, NLSE takes the following dimensionless form. 
\begin{eqnarray*}
i\frac{\partial\psi'}{\partial t'}=-\frac{1}{2}\frac{\partial^2\psi'}{\partial x'^2}+\sqrt{2\pi}\gamma|\psi'|^2\psi'
\end{eqnarray*} 
To avoid notational complication, we will consider all the primed notations to be unprimed. In Fig.\ref{free-dyn-1}, we observe that for $\gamma>0$ the initial wave packet keeps on spreading and try to localize over the space as we predicted analytically.
\begin{figure}[H]
\includegraphics[angle=0,scale=0.8]{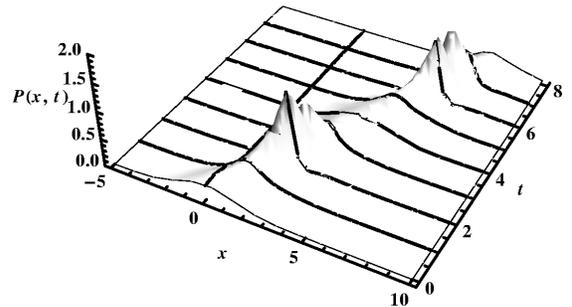}
\caption{Time evolution of initial Gaussian wave packet for $\gamma<0$ for a free particle. $\gamma=-2.0$ and $\delta=2.0$ are considered such that $Y<Y_c$. The width of the packet reduces initially leading to almost collapse at $t=2$.}
\label{free-dyn-2}
\end{figure}
\begin{figure}[H]
\includegraphics[angle=0,scale=0.8]{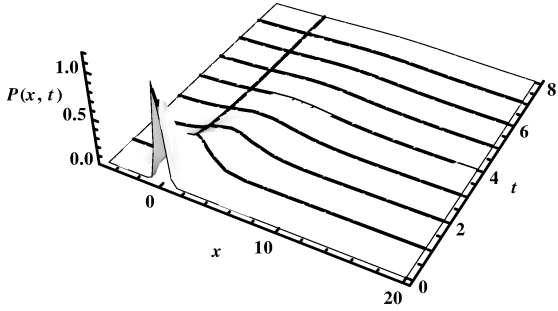}
\caption{Time evolution of initial Gaussian wave packet for $\gamma<0$ for a free particle. $\gamma=-0.5$ and $\delta=0.5$ are considered such that $Y>Y_c$. The initial narrow wave packet keeps on spreading with course of time.}
\label{free-dyn-3}
\end{figure}
\begin{figure}[H]
\includegraphics[angle=0,scale=1.4]{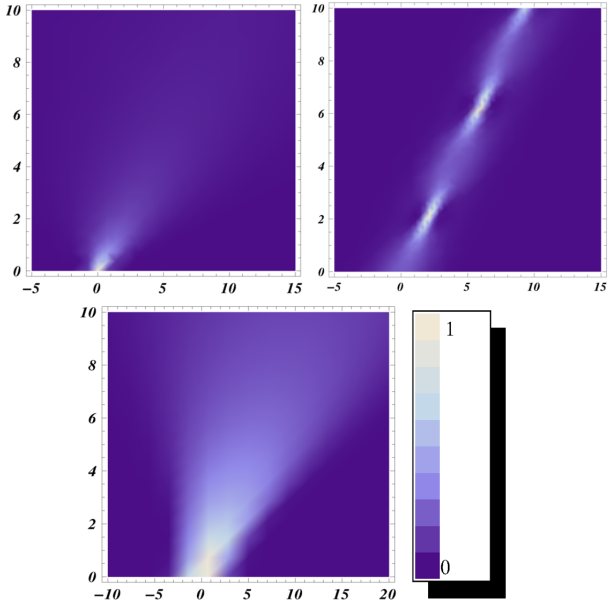}
\caption{Density plot: The dynamics of the free particle wave packet governed by NLSE. Figure at the top-left is for $\gamma<0$ and $\delta=0.5$ such that $Y>Y_c$ and indicates the spreading of initial narrow wave packet. Top-right is for  $\gamma<0$ and $\delta=2.0$ such that $Y<Y_c$ and dictates the possibility of shrinking of the wave packet and collapse in periodic manner for the initial broad wave packet. The figure at the bottom are for $\gamma>0$ and clearly indicates the spreading of the free particle wave packet with time. $x$ and $t$ are considered along X and Y axis respectively.}
\label{free-dyn-4}
\end{figure}
Considering the GPE with harmonic trapping potential given in Eq.(\ref{eq1}) and making the transformations ($t'=\omega t$, $x'=\sqrt{\frac{m\omega}{\hbar}}x$ and $\psi'=(\frac{\hbar}{m\omega})^{1/4}\psi)$ the leading dimensionless form comes out to be
\begin{eqnarray*}
i\frac{\partial\psi'}{\partial t'}=-\frac{1}{2}\frac{\partial^2\psi'}{\partial x'^2}+\sqrt{2\pi}\beta|\psi'|^2\psi'+\frac{1}{2}x'^2
\end{eqnarray*} 
\begin{figure}[H]
\includegraphics[angle=0,scale=0.8]{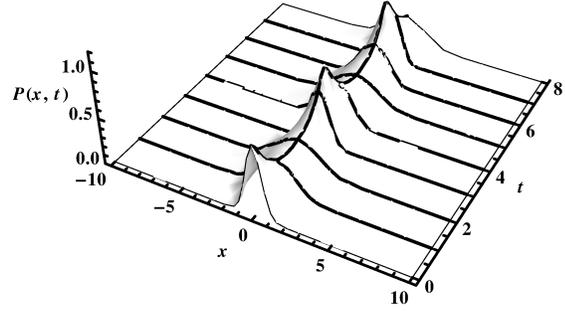}
\caption{Time evolution of initial Gaussian wave packet with SHO potential for $\beta>0$. ($\beta=1.0$ and $\delta=0.75$). The width of initial wave packet oscillates with time.}
\label{sho-dyn-1}
\end{figure}
\begin{figure}[H]
\includegraphics[angle=0,scale=0.8]{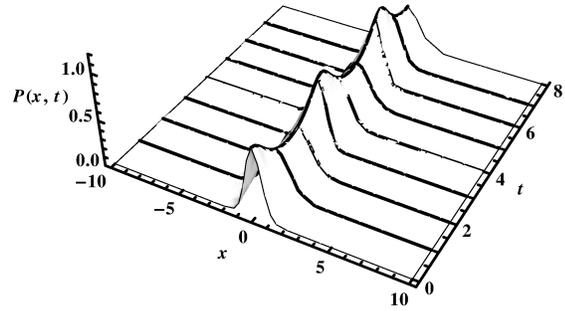}
\caption{Time evolution of initial Gaussian wave packet with SHO potential for $\beta<0$ and when condition given in Eq.(\ref{eq27}) is not satisfied. $\beta=-0.01$ and $\delta=0.75$ are taken . The wave packet width keeps oscillating with time with significant amplitude.}
\label{sho-dyn-2}
\end{figure}
\begin{figure}[H]
\includegraphics[angle=0,scale=0.8]{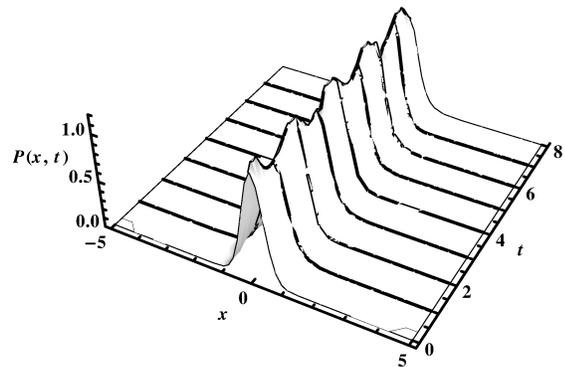}
\caption{Time evolution of initial Gaussian wave packet with SHO potential for $\beta<0$. $\beta=-1.625$ and $\delta=0.5$ are considered such that Eq.(\ref{eq27}) is satisfied, the wave packet width remains almost constant with time.}
\end{figure}
\begin{figure}[H]
\includegraphics[angle=0,scale=1.4]{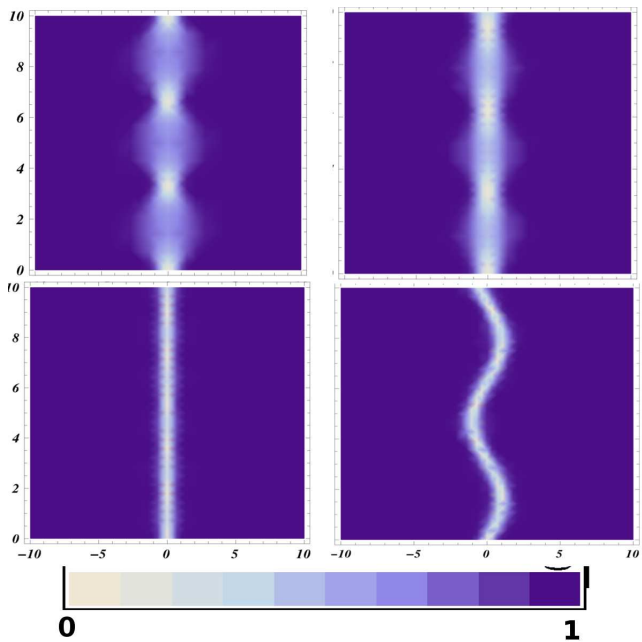}
\caption{Density Plot:The dynamics of the wave packet governed by GPE under harmonic trapping potential. Top row (left to right), are shown for $\beta>0$ and for $\beta<0$ ($\ne\delta(\frac{1}{\delta^2}-3\delta^2)$) respectively. Both the plots at the bottom are for $\beta<0$ and for $\beta=\delta(\frac{1}{\delta^2}-3\delta^2)$). Left is for zero initial momentum where as the right one is plotted for small initial momentum. $x$ and $t$ are considered along X and Y axis respectively.}.
\label{sho-dyn-4}
\end{figure}
\section{Conclusion}
We have shown that that the propagation of wave packet is strongly affected by the nonlinear terms in NLSE. Qualitatively new results are obtained when the coupling constant $g$ is negative. For a harmonic oscillator potential the initial width has to satisfy $\delta^4<\frac{1}{3}$ and the parameter $\beta$ has to choose properly to keep the width unchanged with time (shown in Fig.\ref{fig3}). In Fig.\ref{fig7} we have chosen parameters outside this special range and we have seen the oscillation of the wave packet width with significant amplitude with frequency dictated by the initial width of the wave packet. For the free particle the nonlinearity (if non linear term is negative) has a more interesting impact. If the  width of the wave packet is smaller than the critical value ($\frac{\hbar^2}{m}(\frac{\sqrt{2\pi}}{2|g|})$) then it starts spreading and becomes completely delocalized in space whereas if this width is larger than that critical value then it starts collapsing and ceases to be a Gaussian and develops a secondary maxima. The dynamics of the wave packet has been shown in Fig.\ref{free-dyn-2}.

\section{Acknowledgments}
One of the authors, Sukla Pal would like to thank S. N. Bose National Centre for Basic Sciences for the financial support during the work. Sukla Pal acknowledges Harish-Chandra Research Institute for hospitality and support during visit.

\end{document}